\documentclass[aps,prb]{revtex4}
\usepackage{amsmath}
\usepackage{graphicx}

\begin{document}

\title{Electric
field effect in  heat transfer in 2D devices}

\author{A.I.Volokitin$^{1,2*}$    and B.N.J.Persson$^2$}
\affiliation{
$^1$Samara State Technical University, Molodogvardeiskaya Str.244, 443100 Samara, Russia} \affiliation{$^2$Peter Gr\"unberg Institut,
Forschungszentrum J\"ulich, D-52425, Germany}

\begin{abstract}
We calculate heat transfer between  a 2D sheet (e.g. graphene) and a dielectric in presence of  a gate voltage. The gate potential induces surface charge densities on  the sheet and dielectric, which results in   electric field, which is coupled to  the surface displacements and, as a consequence, resulting an   additional  contributions to the radiative heat transfer. The electrostatic and van der Waals interactions between the surface displacement result in  the   phonon  heat transfer, which we  calculate  taking into account the nonlocality of these interactions. Numerical calculations are presented for heat transfer between graphene and a SiO$_2$ substrate. 
\end{abstract}
\maketitle

PACS: 42.50.Lc, 12.20.Ds, 78.67.-n

\vskip 5mm

\section{Introduction}

The ability to control the electrical properties of materials underlies modern electronics. Currently, in connection with the development of new technologies related to nanoscale thermal management, energy storage and conversion, and information processing, the problem of the management of the heat transfer is being actively considered.
 The problem of heat transfer is, for example, important in the development of a graphene transistor. Heat generation during high-density electric current flowing in graphene will increase the temperature, which can damage the device. Therefore, it is important to be able to control heat flux.

In 2006, Li \textit{et. al.} \cite{Li2006APL} proposed a thermal analogue
of the field effect transistor for controlling heat transfer
by phonons through solid segments, paving the way for
creating blocks for processing information \cite{Li2012RMP,Wang2007PRL} using heat flow instead of electric current. More recently the concept of
thermal transistor has been expanded to noncontact 
systems out of thermal equilibrium \cite{Biehs2014PRL}. In this case, heat fluxes
are associated with the transmission of  thermal photons from one material
to another.

A radiation heat transistor consists of three elements, which by analogy with its electronic counterpart, are called
source, drain and gate. Source and drain are maintained at different temperatures to create a temperature gradient. Source, being traditionally hotter
than drain, emits thermal photons that transmit heat to the drain. These two solids are separated by
intermediate layer made of insulator-metal phase transition material \cite{Biehs2013APL}. This layer acts as a gate.
By adjusting the gate temperature near a critical value,
it is possible radically change the heat flow obtained by
drain and even enhance this flow. The device can work
either at large separation  (far field  \cite{Joulain2015APL}),
where heat fluxes are associated with the propagating  photons,
or at short distances (near field \cite{Biehs2014PRL}), where heat 
is transmitted primarily by photon tunneling. Beyond modulation
and heat flow amplification, these structures
based on phase transition materials can be used for storage
thermal energy and logical operations with thermal photons.

In this article we  consider the possibility of controlling the radiative and phonon heat flux between a 2D sheet (e.g. graphene)  and a dielectric  substrate (e.g. SiO$_2$) using the field effect, similar to how it is used to control electronic properties in  a graphene transistor\cite{Geim2004Science}. The gate voltage induces surface charge densities on  the sheet and dielectric. The thermal fluctuations of the displacements of the charged surfaces give rise to an additional contribution to the fluctuating electromagnetic field which result in  an additional contributions to the radiative heat transfer. The  electrostatic and van der Waals interactions between surface displacements  produce  the  phonon heat transfer.  In contrast to  our previous  ``spring'' model \cite{Persson2011JPCM}, now we take into account the nonlocality in these  interactions. The methodological approach that is used in this article was previously used to calculate heat transfer between 3D metals\cite{Volokitin2019JETPLett}. The electrostatic and van der Waals phonon heat transfer was also recently studied in Ref.\cite{Pendry2016PRB} using different approach.  

\section{Theory}

\subsection{Radiative heat transfer}

Consider a 2D sheet (e.g. graphene) located in the $(x,y)$-plane at $z=0$ and  separated from a dielectric slab with thickness $h$ (e.g. SiO$_2$) by vacuum gap with thickness $d$. The application of the gate voltage (V$_G$) induces a surface charge density on the  sheet, $\sigma_g=ne= E_0/4\pi$, where $n$ is the concentration of the free charge carries in the sheet, $E_0$ is the electric field in the vacuum gap between the sheet and dielectric.  
In general, the application of a gate voltage (V$_G$) creates an electrostatic potential difference $\varphi$  between the graphene and the gate electrode, and the addition of charge carriers leads to a shift in the Fermi level (E$_F$). Therefore $V_G$ is given by \cite{Geim2008NatNanothech}
\begin{equation}
V_G=\frac{E_F}{e}+\varphi
\end{equation}
with $E_F/e$ being determined by the chemical (quantum)
capacitance of the graphene, and $\varphi$ being determined by the
geometrical capacitance $C_G$. 
In this article we consider   the back gate for which\cite{Geim2008NatNanothech}  $V_{BG}\approx \varphi$, where $\varphi=ne/C_G$ is an electrostatic potential difference between the sheet and the gating electrode where $C_G=\varepsilon_d(0)/4\pi h$,  $\varepsilon_d(0)$ is the dielectric constant of the dielectric. The approach which is used in this article can also be applied for the top gate. The radiative heat transfer is associated with the fluctuating electromagnetic field created by
thermal fluctuations of the charge  and current densities inside the bodies. In the case of a charged surface, thermal fluctuations of the surface displacement will also contribute to the fluctuating electromagnetic field and radiative heat transfer. The heat flux between two surfaces separated by a vacuum gap $d$, due to  evanescent (non-radiative) electromagnetic waves (for which $q>\omega/c$)  is determined by the formula
\cite{Volokitin2001PRB,Volokitin2007RMP,Volokitin2017Book}
\begin{equation}
J^{rad} =\int_0^\infty \frac{d\omega}{2\pi}\left[\Pi_g(\omega)-\Pi_d(\omega)\right]\int \frac{d^2q}{(2\pi)^2}
\frac{
4\mathrm{Im}R_{d}(\omega)\mathrm{Im}R_{g}(\omega)e^{-2qd} }{\mid 1-e^{-2
q d}R_{d}(\omega)R_{g}(\omega)\mid ^2},
\label{Heat}
\end{equation}
ãäå
\[
\Pi_{g(d)}(\omega)=\frac{\hbar \omega}{e^{\hbar\omega/k_BT{g(d)}}-1},
\]
 $R_{g(d)}$ is the reflection amplitudes  for the sheet (dielectric) for $p$-polarized electromagnetic waves.

To find the reflection amplitude for the charged sheet,  write the  electric field  of the $p$-polarized electromagnetic wave  in the non-retarded limit in the form: 
\begin{equation}
\mathbf{E}(\mathbf{q}, \omega, z) =e^{i\mathbf{q}\cdot\mathbf{x}-i\omega t}\times \left\{
\begin{array}{rl}
\mathrm{\mathbf{n}}^+e^{-qz}+R_pn^-e^{qz}\mathrm{\mathbf{n}}^-,&  z<0 \\
T\mathrm{\mathbf{n}}^+e^{-qz},& z > 0
\end{array}\right.
\label{wave}
\end{equation}
where $\mathbf{x}=(x,y)$, $\mathbf{q}$ is the wave vector in the  $(x, y)$-plane, $\mathrm{\mathbf{n}}^{\pm}=(\mp i\hat{q}, \hat{z})$. The electric field of the electromagnetic wave induces in the sheet the polarization 
\begin{equation}
\mathbf{p}(\mathbf{x},z) = (p_q\hat{q}+ p_z\hat{z})\delta(z)e^{i\mathbf{q}\cdot\mathbf{x}-i\omega t}
\end{equation}
where $p_q$ and $p_z$ are the parallel and normal components of the surface dipole moment. 
 The boundary condition for the normal component of the electric field  can be obtained from the Maxwell equation
\begin{equation}
\mathbf{\mathrm{\nabla}}\cdot\mathbf{E}=\frac{dE_z}{dz}+iqE_q= -4\pi p_z\delta^{\prime}(z)-iqp_q\delta(z).
\label{maxwell1}
\end{equation}
Integrating  (\ref{maxwell1}) from  $-0$ to $+0$, we get
\begin{equation}
E_z(z=+0)-E_z(z=-0)=-4\pi iqp_q,
\label{gbc1}
\end{equation}
To obtain the boundary condition for the component of the electric field  parallel to the surface, we calculate the circulation of $\mathbf {E}$ around
the contour of the rectangle, two sides of which are parallel to the surface and  located at $ z = \pm 0 $, and the other two sides are perpendicular to the surface. The    magnetic induction field is continuous on the surface, so the magnetic  flux through the area of the rectangle will be zero. Then from the  Faraday's  law  it follows that\cite{Volokitin2019JETPLett,Langreth1989PRB}
\begin{equation}
E_q(z=+0)-E_q(z=-0)=iq\int_{-0}^{+0}dzE_z(z)=-iq\int_{-0}^{+0}dzz\frac{dE_z}{dz}=-4\pi iq p_z.
\label{gbc2}
\end{equation}
The electromagnetic wave will create mechanical stress on the sheet $\tilde{\sigma}_{zi}=\sigma_{zi}(z=+0)-\sigma_{zi}(z=-0)$, where $\sigma_{zi}(z=\pm 0)$ is the Maxwell stress tensor at $z=\pm 0$, which in electrostatic
limit has the form
\begin{equation}
\sigma_{ij}=\frac{1}{4\pi}\left(\tilde{E}_i\tilde{E}_j-\frac{1}{2}\delta_{ij}\tilde{\mathbf{E}}\cdot\tilde{\mathbf{E}}\right),
\end{equation}
where
\begin{equation}
\tilde{
\mathbf{E}} =\mathbf{E}+\hat{z}\left\{
\begin{array}{rl}
E_0^{\prime},& z > 0\\
-E_0^{\prime},&  z<0
\end{array}\right.
\end{equation}
where $E_0^{\prime}=E_0/2=2\pi \sigma_g $ is the static electric field created by the surface charge density $\sigma_g=en$ on the sheet,  $\mathbf{E}$ is the electric field  of the electromagnetic wave, which is determined by  ($\ref{wave}$).   To linear order in the amplitude of the electromagnetic wave, the perpendicular component of the stress   is  determined by 
\begin{equation}
\tilde{\sigma}_{zz}\approx \sigma_g \frac{E_z(+0)+E_z(-0)}{2},
 \label{sigperp}
\end{equation}
where $E_{z(q)}^{\pm}=E_{z(q)}(z=\pm 0)$. The stress $\tilde{\sigma}_{zz}$ produces  a polarization of the sheet,  with normal component  of the dipole moment 
\begin{equation}
p_{z}=\sigma_g u_z=\sigma_g^2M_g\frac{E_z(+0)+E_z(-0)}{2},
\end{equation}
where  $M_g$ is the mechanical susceptibility  of the sheet which determines the surface displacement of the sheet  under an action of the stress $\tilde{\sigma}_{zz}$: $u_z=M_g\tilde{\sigma}_{zz}$.  
The stress  $\tilde{\sigma}_{zq}$, which produces the polarization parallel to the surface  of the sheet, is 
\begin{equation}
\tilde{\sigma}_{zq}\approx ne\frac{E_q^++E_q^-}{2}
\label{parpolarization}
\end{equation}
Thus, the polarization of graphene parallel to the surface arises due to the effective electric field $E_{\parallel}^{\mathrm{eff}}=(E_q^++E_q^-)/2$,
which induces   the sheet dipole moment $p_{q}=i\sigma (E_q^++E_q^-)/2\omega$  parallel to the surface, where $\sigma$ is the sheet conductivity. The sheet conductivity can be expressed in terms of the dielectric function $\varepsilon_g = 1+2\pi iq\sigma/\omega$.
From  (\ref{wave}), (\ref{gbc1}) and (\ref{gbc2}), the boundary conditions can be written as
\begin{equation}
T-1-R=-(\varepsilon_g-1)(T+1-R),
\label{gbc3}
\end{equation}
\begin{equation}
T-1+R=2\pi q\sigma_g^2M_g(T+1+R),
\label{gbc4}
\end{equation}
From  (\ref{gbc3}) and  (\ref{gbc4}) follows
\begin{equation}
R_g=\frac{\varepsilon_g -1 +2\pi q\sigma_g^2M_g}{\varepsilon_g(1  -2\pi q\sigma_g^2M_g)}.
\label{rcg1}
\end{equation}
The reflection amplitude (\ref{rcg1}) has the resonances at $\mathrm{Re}\varepsilon_g=0$ and $1  -2\pi q\sigma_g^2\mathrm{Re}M_g=0$ associated with the plasmon and phonon polaritons of the sheet, respectively. Close to the phonon polariton resonance 
\begin{equation}
R_g\approx\frac{1}{1  -2\pi q\sigma_g^2M_g}.
\label{rcgph}
\end{equation} 
Thus in this case the reflection amplitude depends only on $M_g$ and does not depend on $\varepsilon_g$ what means that at the phonon polariton resonance the optical properties of the sheet are determined only by the mechanical properties of the flexural mode. The reflection amplitude  for the charged dielectric surface  can be calculated in the same way as above for the sheet. Neglecting the polarization  parallel to the surface, the boundary conditions on the surface of the dielectric can be written as\begin{equation}
E_z(z=-d+0)=\varepsilon_d(\omega)E_z(z=-d-0),
\label{bcd1}
\end{equation}
\begin{equation}
E_q(z=-d+0)-E_q(z=-d-0)=-4\pi iqp_z^d,
\label{bcd2}
\end{equation}
where $\mathbf{E}(z>-d)$ determines the incident and reflected wave, ànd  $\mathbf{E}(z<-d)$ is the refracted wave.
For a constant electric field, the surface density of the polarization charge for the dielectric
\begin{equation}
\sigma_d=-\frac{\varepsilon_d(0)-1}{\varepsilon_d(0)}\sigma_g,
\label{sprime}
\end{equation}
where  $\epsilon _{d}(\omega)$ is the dielectric function of a substrate, and $p_z^d= \sigma_dM_dE_z(d=-d+0)$ is the dipole moment normal to the surface. 
From  (\ref{bcd1}) è (\ref{bcd2}) the reflection amplitude for a charged dielectric surface is 
\begin{equation}
R_d=\frac{\varepsilon_d -1 +4\pi q\sigma_d^{ 2}M_d\varepsilon_d}{\varepsilon_d+1  -4\pi q\sigma_d^{2}M_d\varepsilon_d}.
\label{rcd1}
\end{equation}
 For metals $\sigma_g=-\sigma_d$, and in this case (\ref{rcd1}) is reduced to formula obtained in Ref.\cite{Volokitin2019JETPLett}. The reflection amplitude has resonance at
\begin{equation}
\mathrm{Re}[\varepsilon_d(1-4\pi q\sigma_d^{2}M_d)]+1=0.
\label{refcd}
\end{equation}
For a polar dielectric, this resonance is associated with surface phonon polaritons arising from the hybridization of optical and acoustic waves.  
For small potential difference,  when $2\pi q\sigma_g^2M_g\ll 1$ and $4\pi q\sigma_d^2M_d\ll 1$, the reflection amplitudes are determined by well known formulas \cite{Volokitin2017Book}
\begin{equation}
R_g=\frac{\varepsilon_g -1 }{\varepsilon_g },
\label{rcg2}
\end{equation}
\begin{equation}
R_d=\frac{\varepsilon_d -1 }{\varepsilon_d +1}.
\label{rcd2}
\end{equation}

\subsection{Phonon heat transfer}

\subsubsection{van der Waals interaction between semi-infinite medium and a 2D sheet}

In the case of the van der Waals interaction between a semi-infinite dielectric  and a 2D sheet, 
such as graphene,
the potential energy can be written as

\[
U =C_2\int d^2\mathbf{x}_1\int d^2\mathbf{x}_2\int_{-\infty}^{u_d(\mathbf{x}_1)}dz_1
\frac{1}{[(\mathbf{x}_1-\mathbf{x}_2)^2+
(d+u_g(\mathbf{x}_2)-z_1)^2]^6}
\]
\begin{equation}
-C_1\int d^2\mathbf{x}_1\int d^2\mathbf{x}_2
\int_{-\infty}^{u_d(\mathbf{x}_1)}dz_1\frac{1}{[(\mathbf{x}_1-\mathbf{x}_2)^2+
(d+u_g(\mathbf{x}_2)-z_1)^2]^3},
\label{2Duvdw}
\end{equation}
where $u_d (\mathbf{x})$ and $u_g(\mathbf{x})$ are the surface displacements for the dielectric and sheet. Expanding (\ref{2Duvdw}) to second order in displacements, we get
\[
U =\pi A\left(\frac{C_2}{45d^9}-\frac{C_1}{6d^3}\right)
+\pi\left[\int d^2\mathbf{x}_1u_d(\mathbf{x}_1)-
\int d^2\mathbf{x}_2u_g(\mathbf{x}_2)\right]\left(\frac{ C_2}{5d^{10}}-\frac{ C_1}{2d^4}\right)
\]
\[
+6d \int d^2\mathbf{x}_1\int d^2\mathbf{x}_2u_d(\mathbf{x}_1)u_g(\mathbf{x}_2)
\left\{\frac{C_1}{[(\mathbf{x}_1-\mathbf{x}_2)^2+
d^2]^4}-\frac{2C_2}{[(\mathbf{x}_1-\mathbf{x}_2)^2+
d^2]^7}\right\}
\]
\begin{equation}
+\pi\left(\frac{ C_2}{ d^{11}}-\frac{ C_1}{ d^5}\right)\left[\int  d^2\mathbf{x}_1u_d^2(\mathbf{x}_1)+
\int  d^2\mathbf{x}_2u_2^2(\mathbf{x}_2)\right]+...,
\label{uappr}
\end{equation}
At the equilibrium distance the linear terms in the displacement   vanish. Thus, the equilibrium 
distance  $d_0=(2C_2/5C_1)^{1/6}$. The stresses that act on  surfaces of the dielectric  and sheet,  when they are displaced, are 
determined by  
\begin{equation}
\sigma_d=-\frac{\delta U}{\delta u_d(\mathbf{x})}=-6d C_1\int d^2\mathbf{x}_2u_g(\mathbf{x}_2)
\left\{\frac{1}{[(\mathbf{x}-\mathbf{x}_2)^2+
d^2]^4}-\frac{5d_0^6}{[(\mathbf{x}-\mathbf{x}_2)^2+
d^2]^7}\right\}
-\frac{ \pi C_1u_d(\mathbf{x})}{ d^{5}}\left[5\left(\frac{d_0}{d}\right)^6-2\right],
\label{sigmad}
\end{equation}
\begin{equation}
\sigma_g=-\frac{\delta U}{\delta u_g(\mathbf{x})}=-6d C_1\int d^2\mathbf{x}_2u_d(\mathbf{x}_2)
\left\{\frac{1}{[(\mathbf{x}-\mathbf{x}_2)^2+
d^2]^4}-\frac{5d_0^6}{[(\mathbf{x}-\mathbf{x}_2)^2+
d^2]^7}\right\}
-\frac{ \pi C_1u_g(\mathbf{x})}{ d^{5}}\left[5\left(\frac{d_0}{d}\right)^6-2\right].
\label{sigma2}
\end{equation}
Using  a Fourier transformation
\begin{equation}
u_i(\mathbf{x})=\int \frac{d^2\mathbf{q}}{(2\pi)^2}u_ie^{i\mathbf{q}\cdot \mathbf{x}},
\end{equation}
we get
\begin{equation}
\sigma_d=au_d-bu_g,
\label{vdwsigmad}
\end{equation}
\begin{equation}
\sigma_g=au_g-bu_d,
\label{vdwsigmag}
\end{equation}
where
\begin{equation}
a=\frac{\pi C_1}{d^5}\left[2-5\left(\frac{d_0}{d}\right)^6\right],
\label{avdw}
\end{equation}
\begin{equation}
b=\frac{\pi q^3C_1}{4d^2}\left[K_3(qd)-\frac{q^3d ^6_0K_6(qd)}{192d^3}\right],
\label{bvdw}
\end{equation}
where we have used that
\begin{equation}
\int d^2\mathbf{x}_1\frac{u_1(\mathbf{x}_1)}{[(\mathbf{x}_1-\mathbf{x})^2+
d^2]^{\mu +1}}=\int \frac{d^2\mathbf{q}}{(2\pi)^2}u_ie^{i\mathbf{q}\cdot \mathbf{x}}G_{\mu +1}(q),
\end{equation}
\begin{equation}
G_{\mu +1}(q)=\int d^2\mathbf{x}\frac{e^{i\mathbf{q}\cdot\mathbf{x}}}{(r^2+
d^2)^{\mu +1}}=2\pi\int_0^{\infty}\frac{J_0(qr)r dr}{(r^2+d^2)^{\mu+1}}=\frac{\pi}{2^{\mu-1}} 
\left(\frac{q}{d}\right)^{\mu}\frac{K_{\mu}(qd)}{\Gamma(\mu +1)}
\end{equation}
where $K_{\mu}(z)$ is the Bessel function of the second kind and order $\mu$ (see Ref.\cite{Handbook}). For small argument the Bessel function can be approximated by formula\cite{Handbook} 
\begin{equation}
K_{\nu}(z)\sim \frac{2^{\nu -1}\Gamma(\nu)}{z^{\nu}}
\label{besappr}
\end{equation}
where $z\ll 1$. 
Using Eq.(\ref{besappr}) it can be shown that for $qd\ll 1$ the ``spring'' model\cite{Persson2011JPCM} is valid for which $b=a=-K$, where $K$ is a spring constant per unit area characterizing the interaction between the two solids\cite{Persson2011JPCM}. At the equilibrium distance 
\begin{equation}
a=-\frac{3\pi C_1}{d_0^5}=-K_0,
\label{avdw}
\end{equation}
where $K_0$ is the spring constant at $d=d_0$. 
Thus the parameters of the interaction can be written in the form 
\begin{equation}
a=\frac{K_0}{3}\left(\frac{d_0}{d}\right)^5\left[2-5\left(\frac{d_0}{d}\right)^6\right],
\label{2avdw}
\end{equation}
\begin{equation}
b=\frac{q^3d_0^5K_0}{12d^2}\left[K_3(qd)-\frac{q^3d ^6_0K_6(qd)}{192d^3}\right].
\label{2bvdw}
\end{equation}

\subsubsection{Electrostatic interaction between dielectric surface and conducting 2D sheet}
  
The electrostatic potential in the vacuum gap between the 2D sheet and dielectric has the form
\begin{equation}
\varphi = \sigma_g z +\mathrm{const}.
\end{equation}
The surface displacements of the sheet and dielectric
\begin{equation}
u_{zi}(\mathbf{x})=u_ie^{i\mathbf{q}\cdot \mathbf{x}}
\end{equation}
will give rise to normal component of surface dipole moments $p_{zi}=\sigma_iu_i$ resulting in  a change of the electric field in the vacuum gap, which can be described by the potential  
\begin{equation}
\phi(\mathbf{x},z)=\left(\nu_- e^{-qz}+\nu_+ e^{qz}\right)e^{i\mathbf{q}\cdot\mathbf{x}}.
\end{equation}
According to  Eqs. (\ref{bcd1})-(\ref{sprime}), on the dielectric surface the boundary condition for the electric field  takes the form 
\begin{equation}
\nu_-e^{qd}= -R_{d0}\nu_+e^{-qd}-4\pi \sigma_gu_dR_{d0},
\label{elbcd2}
\end{equation}
where $R_{d0}$ is determined by (\ref{rcd2}) at $\omega=0$.
In the electrostatic limit, the sheet potential should remain unchanged when the surface is displaced, from which follows the boundary condition\cite{Pendry2016PRB} 
\begin{equation}
4\pi \sigma_gu_g+\nu_- +\nu_+ =0.
\label{elbcg}
\end{equation}
This Eq. (\ref{elbcg}) also can be obtained from  (\ref{elbcd2}) with $R_{d0}=1$, $d=0$ and changing $u_d$ to $u_g$. 

From  (\ref{elbcg}) and (\ref{elbcd2}) 
\begin{equation}
\nu_+ =\frac{E_0}{1-e^{-2qd}R_{d0}}\left(e^{-qd}R_{d0}u_d-u_g\right),
\label{nu+}
\end{equation}
\begin{equation}
\nu_- =\frac{e^{-qd}R_{d0}E_0}{1-e^{-2qd}R_{d0}}\left(e^{-qd}u_g-u_d\right).
\label{nu-}
\end{equation}
The electric field normal to the surfaces of the 2D sheet and dielectric for  $-d<z<0$ is
\begin{equation}
E_z =-E_0+\frac{qE_0}{1-e^{-2qd}R_d^0}\left[\left(e^{qz}+e^{-2qd}R_{d0}e^{-qz}\right)u_g-e^{-qd}\left(e^{qz}+e^{-qz}\right)R_{d0}u_d\right]e^{\mathbf{q}\cdot \mathbf{x}},
\label{efg}
\end{equation}
and  from Maxwell stress tensor, the stresses that act on the surfaces of the sheet  and dielectric  due to the surface displacements are
\begin{equation}
\sigma_g=K_gu_g-Ku_d,
\label{sigmag}
\end{equation}
\begin{equation}
\sigma_d=K_{g}u_d-Ku_g,
\label{sigmad}
\end{equation}
where
\begin{equation}
K_{g}=\frac{E_0^2}{4\pi}\frac{q\left(1+e^{-2qd}R_{d0}\right)}{1-e^{-2qd}R_{d0}},
\end{equation}
\begin{equation}
K_{d}=\frac{E_0^2}{4\pi}\frac{qR_{d0}\left(1+e^{-2qd}\right)}{1-e^{-2qd}R_{d0}},
\end{equation}
\begin{equation}
K=\frac{E_0^2}{2\pi}\frac{qe^{-qd}R_{d0}}{1-e^{-2qd}R_{d0}}.
\end{equation}

\subsubsection{Phonon heat flux between surfaces}

The displacements  of the surfaces $u_i = u_i^f
+u_i^{ind}$, where $u^f_i$ is the fluctuating displacement due to thermal and quantum
fluctuations inside the bodies and $u^{ind}_i$ is the induced displacement, which occurs due to the interaction between surfaces. Thus, according to Eqs. (\ref{vdwsigmad}), (\ref{vdwsigmag}),(\ref{sigmag}) and (\ref{sigmad}) the displacements of the surfaces are determined by
\begin{equation}
u_g= u_g^f+M_g\sigma_g=u_g^f+M_g(K_gu_g-Ku_d),
\label{eqg}
\end{equation}
\begin{equation}
u_d= u_d^f+M_d\sigma_d=u_d^f+M_d(K_du_d-Ku_g),
\label{eqd}
\end{equation}
where $M_i$ is the mechanical susceptibility which determines the surface displacement under an action of the applied stress: $u_i^{ind}=M_i\sigma_i$, and where 
\begin{equation}
K_{g}=\frac{E_0^2}{4\pi}\frac{q\left(1+e^{-2qd}R_{d0}\right)}{1-e^{-2qd}R_{d0}}+\frac{K_0}{3}\left(\frac{d_0}{d}\right)^5\left[2-5\left(\frac{d_0}{d}\right)^6\right],
\end{equation}
\begin{equation}
K_{d}=\frac{E_0^2}{4\pi}\frac{qR_{d0}\left(1+e^{-2qd}\right)}{1-e^{-2qd}R_{d0}}+\frac{K_0}{3}\left(\frac{d_0}{d}\right)^5\left[2-5\left(\frac{d_0}{d}\right)^6\right],
\end{equation}
\begin{equation}
K=\frac{E_0^2}{2\pi}\frac{qe^{-qd}R_{d0}}{1-e^{-2qd}R_{d0}}+\frac{q^3d_0^5K_0}{12d^2}\left[K_3(qd)-\frac{q^3d ^6_0K_6(qd)}{192d^3}\right].
\end{equation}

According to the fluctuation-dissipation theorem, the spectral density of fluctuations of the surface displacements
is determined by \cite{LandauStatisticalPhysics}
\begin{equation}
\langle|u_{g(d)}^f|^2\rangle = \hbar \mathrm{Im}M_{g(d)}(\omega,q)\coth\frac{\hbar\omega}{2k_BT_{g(d)}}.
\label{fdt}
\end{equation}
From (\ref{eqg}) and (\ref{eqd}) 
\begin{equation}
u_g=\frac{(1-M_dK_d)u_g^f-M_gKu_d^f}{(1-M_gK_g)(1-M_dK_d)-K^2M_gM_d},
\label{ugq}
\end{equation}
$u_d$ is obtained from $u_g$ by the permutation  of indexes ($g \leftrightarrow d$).
The mechanical stress that acts on the surface of the dielectric,  due to fluctuations of the surface displacement of     the sheet,  is determined by
\begin{equation}
\sigma_{dg}^f=\frac{Ku_g^f}{(1-M_gK_g)(1-M_dK_d)-K^2M_gM_d}.
\label{sigmsdf}
\end{equation}
The stress $\sigma_{gd}^f$ can be obtained from $\sigma_{dg}^f$
by the permutation ($ g \leftrightarrow d $).
The   heat flux from the sheet  to dielectric due to the electrostatic and van der Waals interactions between the fluctuating surface displacements  is   \cite{LandauStatisticalPhysics,Volokitin2019JETPLett} 
\[
J^{ph}=\langle\dot{u}_d^{ind}\sigma_{dg}^f\rangle-\langle\dot{u}_g^{ind}\sigma_{gd}^f\rangle=
2\int_0^\infty\frac{d\omega}{2\pi}\int\frac{d^2q}{(2\pi)^2}\omega\left[\mathrm{Im}M_d\langle|\sigma_{dg}|^2\rangle-
\mathrm{Im}M_g\langle|\sigma_{gd}|^2\rangle\right]=
\]
\begin{equation}
=\frac{1}{\pi^2}\int_0^\infty d\omega\left[\Pi_g(\omega)-\Pi_d(\omega)\right]\int_0^\infty dqq
\frac{
K^2\mathrm{Im}M_g\mathrm{Im}M_d }{\mid (1-K_gM_g)(1-K_dM_d)-
K^2M_gM_d\mid ^2}.
\label{heatflux}
\end{equation}
In the ``spring'' model\cite{Persson2011JPCM} it is assumed that $K_g=K_d=-K$ and in this case  (\ref{heatflux}) is reduced to 
\begin{equation}
J^{ph}_{spr}=\frac{1}{\pi^2}\int_0^\infty d\omega\left[\Pi_g(\omega)-\Pi_d(\omega)\right]\int_0^\infty dqq
\frac{
K^2\mathrm{Im}M_g\mathrm{Im}M_d }{\mid 1+K(M_g+M_d)\mid ^2}.
\label{heatfluxspr}
\end{equation}

\section{Numerical results for graphene on a SiO$_2$ substrate}

\begin{figure}
\includegraphics[width=0.5\textwidth]{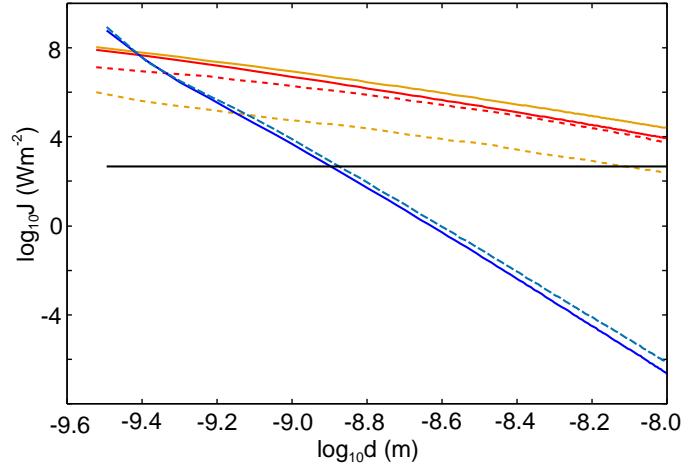}
\caption{ Dependence of the heat flux between graphene and a SiO$_2$ substrate   on the separation between them for different mechanisms. The red and brown 
 lines show the heat flux for the radiative  and electrostatic phonon  heat transfer: the solid and dashed  lines for the concentration of the free charge carries in graphene $n_g=10^{19}$m$^{-2}$ and $n_g=10^{18}$m$^{-2}$, respectively. The blue lines show the heat flux associated with   the van der Waals interaction: the solid and dashed lines are for the nonlocal and local (the ``spring'' model) theories of the van der Waals phonon  heat transfer.   The black line shows the
radiative heat transfer associated with blackbody radiation.   \label{Distance}}
\end{figure}

For an elastic sheet\cite{Landau1970}
\begin{equation}
M_g=\frac{1}{\kappa q^4-\rho \omega^2-i\omega \rho\gamma},
\end{equation}
where the bending stiffness of graphene    $\kappa\approx 1$eV, $\rho=7.7\cdot 10^{-7}$kg/m$^2$ is the surface mass density of   graphene, $\gamma$ is the damping constant for flexural motion of graphene which was estimated in Ref.\cite{KapitzRes2016PRB} as
\begin{equation}
\gamma=\frac{\omega T}{100T_{RT}}
\end{equation}
where $T_{RT}=300$K is the room temperature.

For an elastic semi-infinite medium the mechanical susceptibility   $M_d$ is determined by formula \cite{Persson2001JPCM}
\begin{equation}
M_d=\frac{i}{\rho c_t^2}\left(\frac{\omega}{c_t}\right)^2\frac{p_l(q,\omega)}{S(q,\omega)},
\end{equation}
where
\[
S(q,\omega)=\left[\left(\frac{\omega}{c_t}\right)^2-2q^2\right]^2+4q^2p_tp_l,
\]
\[
p_t=\left[\left(\frac{\omega}{c_t}\right)^2-q^2+i0\right]^{1/2}, \,\,p_l=\left[\left(\frac{\omega}{c_l}\right)^2-q^2+i0\right]^{1/2},
\]
where for SiO$_2$ the density  $\rho=2.65\cdot 10^3$kg/m$^{3}$,  $c_l=5968$m/s and $c_t=3764$m/s  are the longitudinal and transverse velocities of the acoustic waves.

The dielectric function of amorphous SiO$_2$ can be described
using an oscillator model\cite{Chen2007APL}
\begin{equation}
\varepsilon(\omega) =
\epsilon_{\infty}+\sum_{j=1}^2\frac{\sigma_j}{\omega_{0,j}^2-\omega^2-i\omega\gamma_j},
\end{equation}
where parameters $\omega_{0,j}$, $\gamma_j$ and $\sigma_j$ were
obtained by fitting the measured $\varepsilon$ for SiO$_2$ to the above
equation, and are given by $\epsilon_{\infty}=2.0014$,
$\sigma_1=4.4767\times10^{27}$s$^{-2}$, $\omega_{0,1}=8.6732\times
10^{13}$s$^{-1}$, $\gamma_1=3.3026\times 10^{12}$s$^{-1}$,
$\sigma_2=2.3584\times10^{28}$s$^{-2}$, $\omega_{0,2}=2.0219\times
10^{14}$s$^{-1}$, and $\gamma_2=8.3983\times 10^{12}$s$^{-1}$.

\begin{figure}
\includegraphics[width=0.5\textwidth]{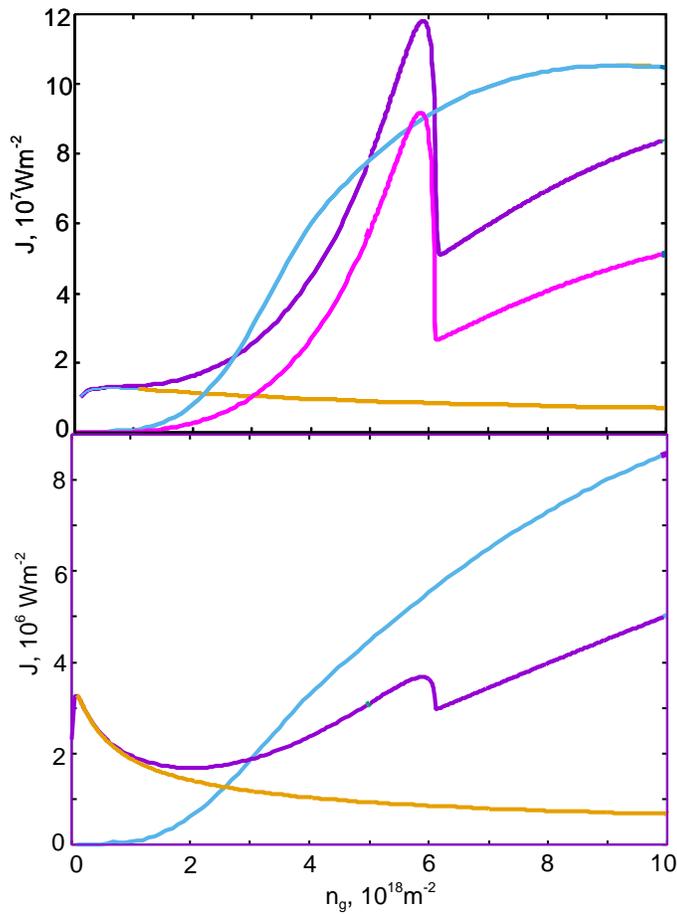}
\caption{ Dependence of the heat flux between graphene and a SiO$_2$ substrate   on the concentration of the free charge carries in graphene. The violet and blue lines show  the radiative and  electrostatic phonon  heat flux at $d=0.3$nm (top) and $d=1$nm (bottom). The brown lines  show  the radiative heat flux without contribution from the thermally fluctuating surface displacements, while the pink line on the top shows only these contributions.  \label{Density}}
\end{figure}

In the numerical calculations we used the dielectric function of graphene, which was
calculated  within the random-phase approximation (RPA)
\cite{Wunsch2006NJP,Hwang2007PRB}.  The dielectric function
is an analytical function in the upper half-space
of the complex $\omega$-plane:
\begin{equation}
\varepsilon_g(\omega,q)=1+\frac{4k_Fe^2}{\hbar
v_Fq}-\frac{e^2q}{2\hbar \sqrt{\omega^2-v_F^2q^2}}\Bigg \{G\Bigg
(\frac{\omega+2v_Fk_F}{v_Fq}\Bigg )- G\Bigg
(\frac{\omega-2v_Fk_F}{v_Fq}\Bigg )-i\pi \Bigg \},
\end{equation}
where
\begin{equation}
G(x)=x\sqrt{x^2-1} - \ln(x+\sqrt{x^2-1}),
\end{equation}
where the Fermi wave vector $k_F=(\pi n)^{1/2}$, $n$ is the
concentration of charge carriers, the Fermi energy
$\epsilon_F=\hbar v_Fk_F$,  $v_F\approx 10^6$ m/s is the Fermi velocity.

For graphene on a SiO$_2$ substrate at equilibrium distance, according to functional theory calculation, the spring constants\cite{KapitzRes2016PRB} $K_0=K_{OH}=1.23\cdot 10^{20}$N/m$^3$ and $K_0=K_{H}=1.56\cdot 10^{20}$N/m$^3$ for the OH- and H-terminated SiO$_2$ substrate, which agrees rather well with estimation $K_0=1.82\cdot10^{20}$N/m$^3$ in Ref.\cite{Persson2011JPCM}

Fig. \ref{Distance} shows the dependence of the heat flux between graphene and a SiO$_2$ substrate   on the separation between them for different mechanisms and different free  charge carries concentration $n$. The heat fluxes associated with  the radiative and electrostatic phonon heat transfer have practically the same distance dependence and  the difference between them decreases when the concentration $n$ increases. The heat flux associated with the van der Waals interaction decays with distance much faster than  for the radiative and electrostatic phonon heat transfer. Thus for $n_g>10^{19}$m$^{-2}$ the radiative and electrostatic phonon heat transfer dominate for practically  all separations. The blue solid and dashed lines show the contributions to the heat flux calculated using nonlocal  and local theories of the van der Waals phonon heat transfer. For distances $d<10$nm the difference between these contributions is small, which confirms the validity of the ``spring'' model for such distances. 

Fig. \ref{Density} shows the dependence of the heat flux associated with the radiative and electrostatic phonon heat transfer  between graphene and a SiO$_2$ substrate on the concentration $n_g$ at $d=0.3$nm (top) and $d=1$nm (bottom). For the radiative heat transfer, for $n_g>10^{18}$m$^{-2}$ dominates the contribution which is associated  with flexural vibrational modes of graphene. The sharp maximum in the radiative heat flux at $d=0.3$nm and $n_g\approx 6\cdot 10^{18}$m$^{-2}$ is related with the phonon polariton resonance which occurs due to interaction of the flexural mode with an electric field of the charged sheet. The reflection amplitude for graphene at this resonance is determined by (\ref{rcgph}).

\section{Conclusion}

The heat transfer between a 2D sheet and dielectric in presence of the gate voltage was calculated. The gate voltage  induces a  surface charge density on the  of the sheet and dielectric. As a result, thermal fluctuations of the surface displacements create a fluctuating electromagnetic field that leads to an additional contribution to the radiative heat transfer. The phonon heat transfer due to the electrostatic and van der Waals interactions  between surface displacements was  calculated taking into account the nonlocality of these interactions. Numerical results are presented for the heat transfer between graphene and a SiO$_2$ substrate. It has been shown that, when the charge density $\sigma_g>$0.1Km$^{-2}$ is induced on the graphene surface, the main contribution to the radiative heat transfer is associated with the flexural vibrational modes of graphene, and this contribution is of the same order and has the same distance dependence as the contribution from the electrostatic phonon heat transfer. Due to the strong  distance dependence the van der Waals phonon heat transfer is only important for subnanometer  distances. For $\sigma_g>$1Km$^{-2}$ the heat flux due to  the radiative and electrostatic phonon heat transfer dominate for practically all distances. The obtained results can be important for the heat managements in the 2D devices. 

\vskip 0.5cm

A.I.V. acknowledges  funding by RFBR according to the research project N\textsuperscript{\underline{o}} 19-02-00453. A.I.V. also thanks the Condensed Matter group of ICTP  for hospitality during the time of working on this article.

\vskip 0.5cm

$^*$alevolokitin@yandex.ru

\end{document}